\documentclass[preprint,12pt,3p]{elsarticle}
\usepackage{hyphenat}

\usepackage[
singlelinecheck=false 
]{caption}

\usepackage{enumitem}
\usepackage{bm}
\usepackage{setspace}
\usepackage{makecell}
\usepackage{multirow}

 \usepackage{cancel}
\usepackage{ulem}

\usepackage[dvipsnames,svgnames,x11names]{xcolor}

\usepackage{amsmath}

\usepackage{amssymb}

\journal{Physical Review E}

\begin{document}
\renewcommand{\arraystretch}{1.2}

\begin{frontmatter}

\title{Multirange Ising model on the square lattice}

\author[label1]{Charles S. do Amaral\corref{cor1}}
\address[label1]{Departamento de Matem\'atica - Centro Federal de Educa\c c\~ao Tecnol\'ogica de Minas Gerais, \linebreak Av. Amazonas 7675, Belo Horizonte, MG, Brasil. }

\cortext[cor1]{Corresponding author.}

\ead{charlesmat@cefetmg.br}

\author[label2]{Bernardo N. B. de Lima}
\address[label2]{Departamento de Matem\'atica - Universidade Federal de Minas Gerais.}

\author[label9]{Ronald Dickman}
\address[label9]{Departamento de F\'isica and National Institute of Science and Technology for Complex Systems,\linebreak Universidade Federal de Minas Gerais, Av. Pres. Ant\^onio Carlos, 6627, Belo Horizonte, MG, Brasil.}

\author[label7]{A. P. F. Atman}
\address[label7]{Departamento de F\'isica and National Institute of Science and Technology for Complex Systems, \linebreak Centro Federal de Educação Tecnológica de Minas Gerais.}

\date{\today}

\begin{abstract}

We study the Ising model on the square lattice ($\mathbb{Z}^{2}$) and show, via numerical simulation, that allowing interactions between spins separated by distances $1$ and $m$ (two ranges), the critical temperature, $ T_c (m) $, converges monotonically to the critical temperature of the Ising model on $\mathbb{Z}^4$ as $ m \to \infty $. Only interactions between spins located in directions parallel to each coordinate axis are considered. We also simulated the model with interactions between spins at distances of $ 1 $, $ m $, and $ u $ (three ranges), with $ u $ a multiple of $ m $; in this case our results indicate that $ T_c(m, u) $ converges to the critical temperature of the model on $ \mathbb{Z}^6$. For percolation, analogous results were proven for the critical probability $p_c$ [B. N. B. de Lima, R. P. Sanchis and R. W. C. Silva, Stochastic Process. Appl. {\bf 121}, 2043 (2011)].

\vskip 0.5cm


\end{abstract}

\end{frontmatter}

\section{Introduction}

The Ising model and percolation are among the most important models in statistical mechanics. The former, introduced in $1920$ by Wilhelm Lenz \cite{Lenz}, exhibits a continuous transition between paramagnetic and ferromagnetic phases as temperature $T$ is varied, while the latter, proposed in $ 1957 $ by Broadbent and Hammersley \cite{broadbent} to characterize transport in random media, exhibits a transition between phases with and without global connectivity as the concentration $p$ is varied. A key question regarding these models is the critical value, $T_c$ or $p_c$.
For the $d$-dimensional hypercubic lattice $ \mathbb{Z}^{d}$, the critical value is known exactly only for $ d = 1 $ and $ d = 2 $ \cite{ising, onsager, kesten}. Although quite precise estimates for the critical point are available in some other cases [6-22], the exact values are unknown. 

The present study is motivated by recent work of de Lima, Sanchis and Silva \cite{LSS} as well as previous studies of the Ising model in dimensions $d \geq 4$ [17-22]. The authors of \cite{LSS} consider percolation on $ \mathbb{Z}^d$ adding bonds of $n$ different lengths:

$$ m_1~=~k_1, m_2~=~(k_1 \times k_2), ..., m_n~=~(~k_1~\times~k_2~\times~...~\times k_{n}~)$$

\noindent parallel to each coordinate axis, where $k_i~\in~\{2, 3, ... \}$ for all $i$. These authors prove~that, if $d\geq 2$, the critical point converges to the critical point of percolation on $ \mathbb{Z}^{d (n + 1)}$ as $k_i \to \infty$, for all $i$, in both bond and site percolation. This model is called \textit{multirange percolation}. These authors also conjecture that convergence is monotone and nonincreasing in each variable~$k_i$. Recent numerical work suggests that, if $d=2$, this conjecture is valid and the convergence follows a power law for $ n = 1 $ and $ n = 2 $~\cite{Amaral}. In addition to revealing an unexpected connection between critical values in systems of distinct connectivity, this result allows estimation of the critical point in higher dimensions by simulating the model with multiple ranges in lower dimensions, reducing computational complexity and cost.

These results raise the question of whether other models with local interactions and exhibiting phase transitions  have properties similar to multirange percolation.
In this study, we provide numerical evidence suggesting that this is the case for the Ising model in $d = 2$ dimensions, for $ n = 1 $ (two ranges) or $ n = 2 $ (three ranges). Our results apply to the critical temperature, $T_c $, allowing its determination in higher dimensions by simulating the model with multiple ranges in lower dimensions. 

Turban \cite{turban1} obtained analytical results for an Ising model with $n=1$ in one dimension. He studied a chain of $N$ sites with $ m-spin $ interactions with coupling constant $ J $ in a field $ H $. Using a change of variables, this model can be transformed into the multirange Ising model in $d=1$ with $ n = 1 $ and first-neighbor interactions $ H $ and $ m ^ {th}$-neighbor interactions $ J $. Turban showed that this model can be reinterpreted as a $ 2d $ Ising model in zero external field and with first-neighbor interactions $ H $ and $ J $ (one for each direction) on the rectangular lattice of size $\frac{N}{m}  \times m $. In the thermodynamic limit $ \frac{N}{m} \to \infty $ and $ m \to \infty $, this model displays the critical behavior of the two-dimensional Ising model. Similar results were obtained for the Potts model \cite{turban2}.

The remainder of this paper is organized as follows. Section 2 describes the model and our simulation procedure. Section 3 discusses our results. Key conclusions and open questions are summarized in Section 4.

\section{Model and Numerical Procedure}

To define the multirange model we begin with the usual $d$-dimensional cubic lattice, 
$\mathbb{Z}^d $, and add bonds linking pairs of sites along the principal lattice directions.
The resulting graph $G$ is characterized by $n$ (an integer $\geq 1$) and a set of $n$
integers, $k_1,..., k_n$ (all $\geq 2$), such that the added bonds have length $m_1 = k_1$, $m_2 = k_2 m_1 = k_1 k_2$, and so on, up to $m_n = k_1 \cdots k_n$.  
Thus a site $(x_1, x_2,..., x_d) \in \mathbb{Z}^d $ is connected to its $2d$ nearest neighbors, $(x_1 + 1, x_2, ..., x_d)$, $(x_1 - 1, x_2, ..., x_d)$, ..., $(x_1, ..., x_d -1)$, and in addition to sites at distances $m_1$, $m_2$, ..., $m_n$ along all principal directions.
Fig. 1 shows, for $n=1$ and $m_1 = 2$, a portion of the square lattice $\mathbb{Z}^2$ with the added bonds, highlighted in red (curves), of the central pair of sites.  

With the graph $G$ defined as above, we can implement many multirange statistical models
(spin systems, lattice gases, polymers) by equipping the sites with appropriate variables or operators.  One of the simplest is the multirange Ising model, whose configurational energy (in the absence of an external field) is given by

\begin{figure}[t]
	\begin{center}
		\includegraphics[height=6.0cm]{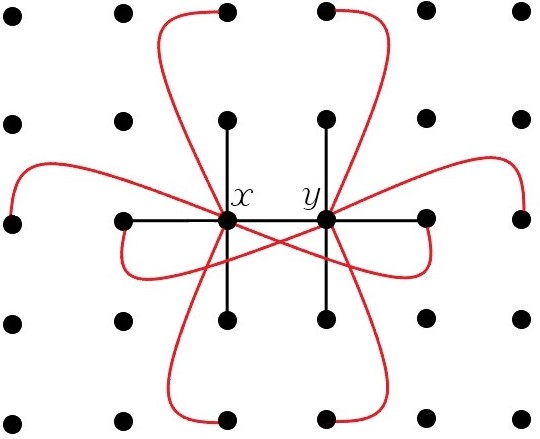} \quad
		\caption{(Color online) Part of the graph with all bonds that terminate at one of the vertices x or y, where $n = 1$ and $m_1 = 2$. The length of bonds with colors black (straight) and red (curved lines) is $1$ and $m_1$, respectively.} \label{fig1}
	\end{center}
\end{figure}

\begin{equation}
\mathcal{H}= -J \displaystyle\sum_{\substack{ (i,j) \in \mathbb{E}_{G}}} \sigma_{i}\sigma_{j}
\end{equation}
\noindent where $\mathbb{E}_{G}$ is the set of bonds of the graph $G$, the spins $\sigma_i$ take values $\{-1, +1\} $ and $J$ is the coupling constant (we take $J = 1$). 

We use the \textit{Wolff algorithm} \cite{wolff} to estimate the Binder cumulant $ U $ and magnetic susceptibility $ \chi $ on multirange graphs constructed on $ L \times L$ square lattices with periodic boundaries.
For $n=1$ (two ranges), we simulate the model with $m_{1}=m$,
where $m \in \{2, 5, 8, 10, 13, 16, 19 \}$; for $n=2$, we use $m_1=m$ and $m_2=m^2$, where $m~\in~\{2, ..., 6\}$. We study system sizes $768 \leq L \leq 2048$ ($n=1$) and $1280 \leq L \leq 2560$ ($n=2$) (note: for $ n = 1$ with $ m = 16$ and $m=19$ , and $n=2$ with $m=6$, we only consider $1024\leq L \leq 2048$ and $ 1536~\leq~L~\leq~2560$, respectively, to reduce finite-size effects). The number of Wolff steps following equilibration ranges from $2 \times 10^6$ ($L=768$) to $10^6$ ($L=2560$). We use the first fifth of the Wolff steps for equilibration.

We estimate the inverse critical temperature, $ K_{c}(m) $, using a procedure
similar to that of \cite{principal}. Initially, we estimate the critical exponent $\nu$ through the relation
\begin{equation}
     \dfrac{\mbox{d}U}{\mbox{d}K}\bigg|_{\mbox{max}} \sim L^{\frac{1}{\nu}} 
\label{nu_gama}
\end{equation}

\noindent where the left-hand side represents the maximum of $\frac{\mbox{d}U}{\mbox{d}K}$ for size $L$.

The effective inverse critical temperature, $ K_{c}(m; L) $, for a system of length $ L $, can be taken as the value that maximizes $\dfrac{\mbox{d}U}{\mbox{d}K}$ or $\chi$. Thus, for each $ m $ and $ L $, we obtain two estimates for $ K_ {c} (m; L) $. Given estimates $K_{c}(m;L)$ for a series of $ L $ values, $K_{c}(m) $ is estimated using the finite-size scaling (FSS) relation,

\begin{equation}
    K_{c}(m) \approx K_{c}(m; L)+\lambda L^{-\frac{1}{\nu}} + \theta L^{-2},
\label{kc}
\end{equation}

\noindent where $\lambda$ and $\theta$ are constants. The correction term $\propto L^{-2}$ is used because without it, the residuals for certain values of $m$ and $n$ exhibit a systematic (parabolic) dependence on $L$. The values obtained for $ \nu $ vary between $0.883(20)$ ($n=1, m_1=19$) and $0.999 (11)$ $(n=1, m_1=2)$. Determination of critical exponents would require a systematic analysis of a larger range of system sizes. In the present context we regard $\nu$ simply as a fitting parameter. 

To estimate $\displaystyle \lim_{m \to \infty} K_c(m)$ we use a three-parameter fit of the form

\begin{equation}
K_{c}(m)=a \, m^{b} + c.
\label{ajuste}
\end{equation}

\noindent Details of the uncertainty analysis are provided in the Appendix.

\begin{figure}[t]
	\begin{center}
		\includegraphics[height=7.9cm]{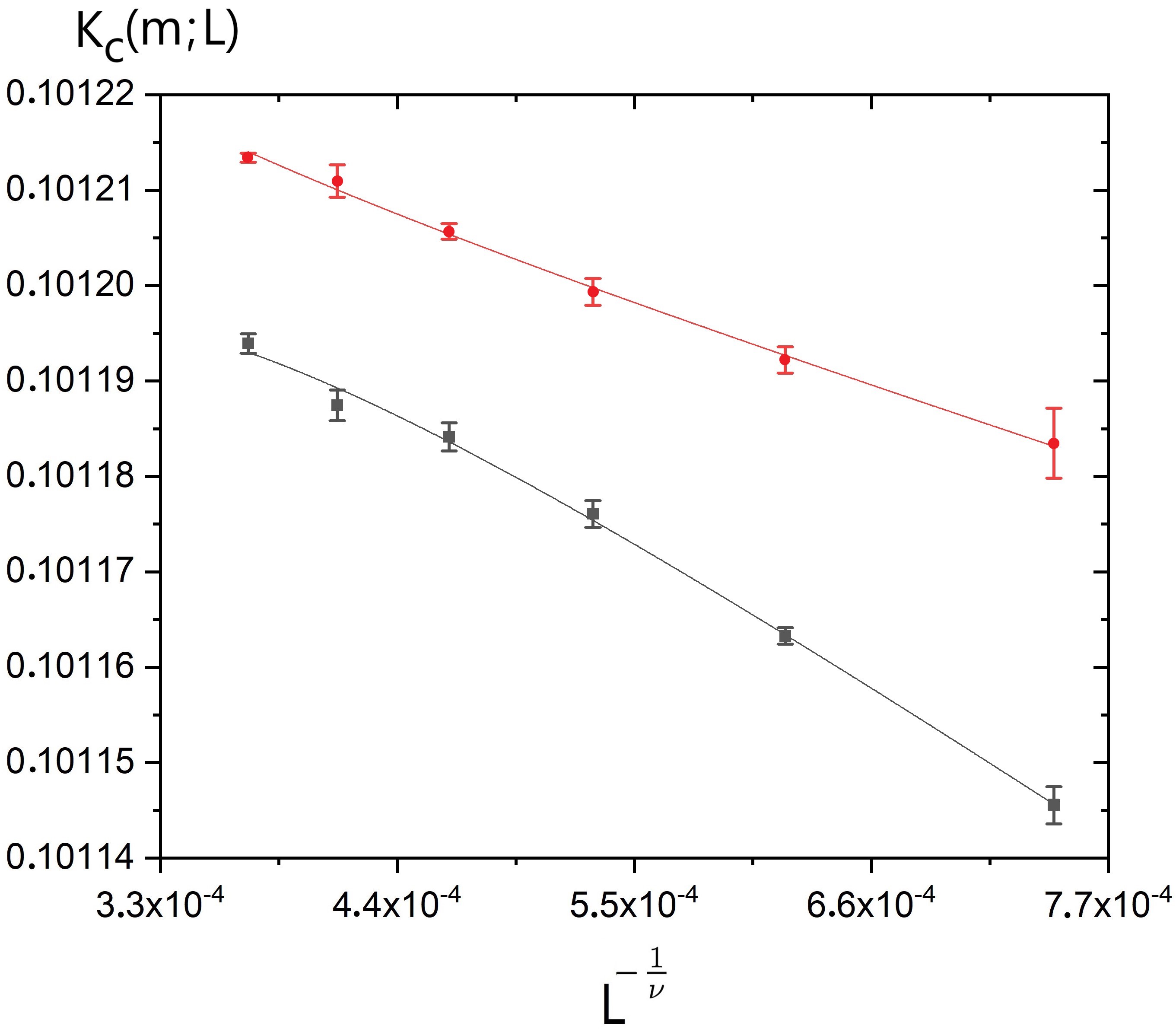}
		\caption{(Color online) Graphs obtained for $ n = 2 $ considering $m=2$ (size ranges: $1$, $2$ and $4$). The plots shows the estimated $ K_{c}(m) $ using the finite-size scaling (3) considering the functions $\frac{\mbox{d}U}{\mbox{d}K}$ (black squares) and $\chi$ (red circles). The lines correspond to fits as described in text.} \label{fig2}
	\end{center}
\end{figure}

\section{Results}

For $ n = 2 $ with $ m = 6 $, Fig. 2 shows plots of the data used to determine $K_c(m)$ through relation (3). The best-fit values of $ a $, $b $, and $ c$ using Eq. (\ref{ajuste}) are summarized in Table 1 and Fig.~3. The data show that $K_c(m)$ is well fit by Eq. (4), and suggest that it converges, as $m \to \infty$, to values close to the inverse critical temperature of the Ising model on $\mathbb{Z}^4 $ (0.1496947(5) \cite{lundow}) for $ n = 1 $, and on $\mathbb{Z}^6$ (0.09229(4) \cite{aktekin1}) for $ n = 2 $.

\begin{table*}[t]
	\centering
	\begin{small}
		\caption*{\ \textbf{Table 1:} Fitting parameters for $K_c (m)$ using
		Eq.~(\ref{ajuste}).}
		\begin{tabular}{c|c|cccc}
			\noalign{\smallskip}
			\hline
			\ $n$ \ \ & \  Function  \ & \ \ \ \ \  $a$  & \ \ \ \ \  $b$ & \ \ \ \ \  $c$ $(K_c)$   & \ \ \ \ \ R-Square \\
			\hline
			\multirow{2}{*}{$1$} \   &  \ $\chi$ & \ \ \ \ \  0.1015(12)  & \ \ \ \ \  -2.070(19)   & \ \ \ \ \  0.149646(46)  & \ \ \ \ \  0.999990\\
			& \ $\frac{\mbox{d}U}{\mbox{d}{K}}$ \ & \ \ \ \ \  0.1007(16)  & \ \ \ \ \  -2.059(24)   & \ \ \ \ \   0.149653(50)  &  \ \ \ \ \  0.999988\\
			\hline
			\multirow{2}{*}{$2$} \ \ &  \ $\chi$\  &  \ \ \ \ \   0.0868(53)     & \ \ \ \ \  -3.258(96)    & \ \ \ \ \  0.092188(78)  & \ \ \ \ \    0.999835 \\
			&  \ $\frac{\mbox{d}U}{\mbox{d}{K}}$  \  &  \ \ \ \ \   0.0847(39)     & \ \ \ \ \  -3.226(74)    & \ \ \ \ \  0.092160(63)  & \ \ \ \ \    0.999934 \\
			\hline
		\end{tabular}
	\end{small}
\end{table*}

 The small discrepancies between our results and literature estimates for $ K_c (\mathbb{Z}^{4})$ and $K_c (\mathbb{Z}^{6})$, obtained using $ \chi $ and $\frac{\mbox{d}U}{\mbox{d}{K}}$, respectively, are likely due to the limited number of Wolff steps employed for each $ L $ value analyzed and/or the limited number of $ m $ values analyzed. Our goal was to be able to study several cases to analyze the behavior of $ K_c (m) $ varying $m$, which required about five months of cpu time on $50$ cores with speed $3.2$ GHz.

 On the basis of the results obtained and the fact that there is analytical proof of convergence of $p_c$ for the multirange percolation model on the square lattice \cite{LSS}, we conjecture that the same holds for the multirange Ising model. All estimated values of $K_{c}(m)$, for $n=1$ and $n=2$, are listed in Table~2.

\begin{figure}[t]
	\begin{center}
		\includegraphics[height=6.2cm]{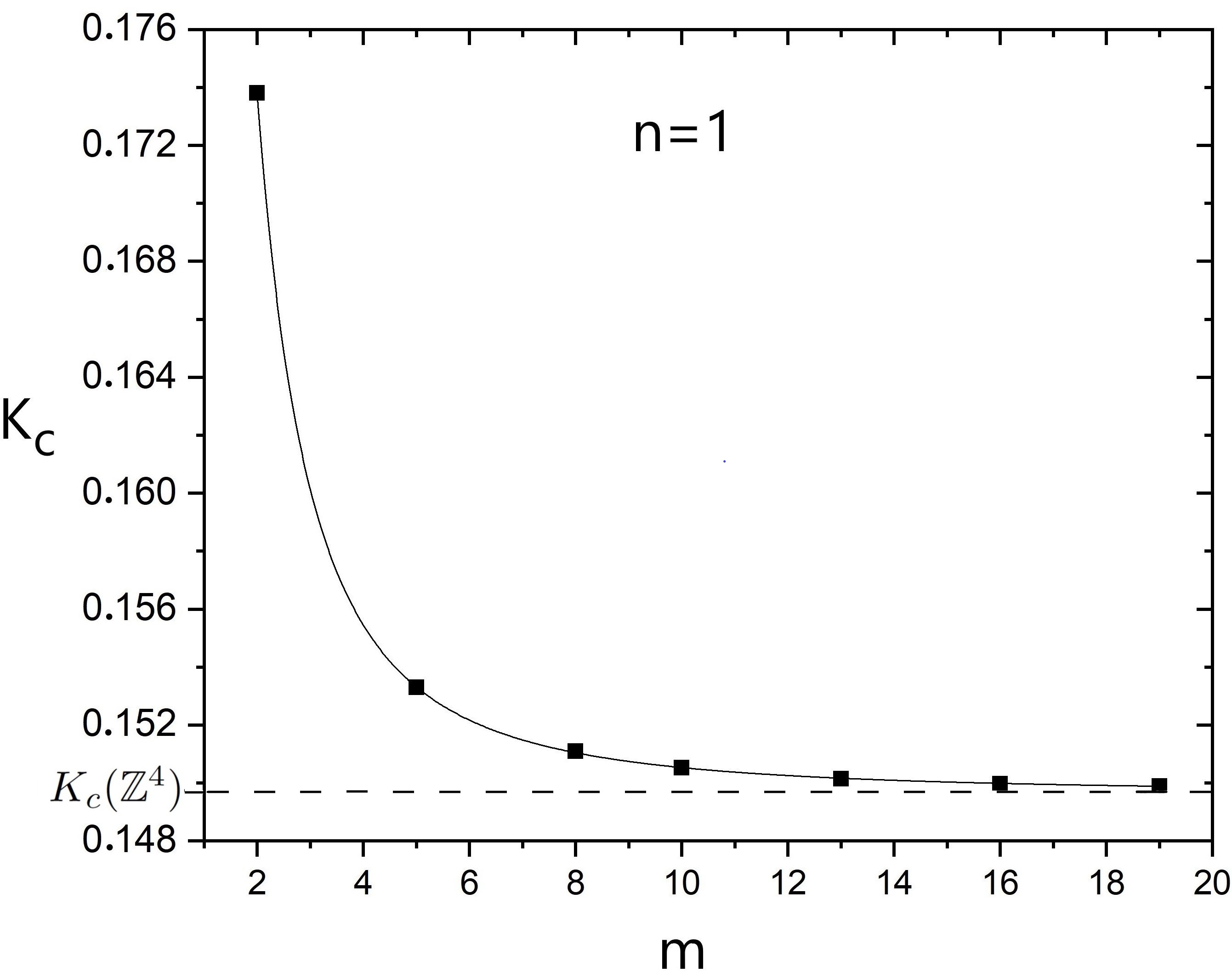} \quad
		\includegraphics[height=6.2cm]{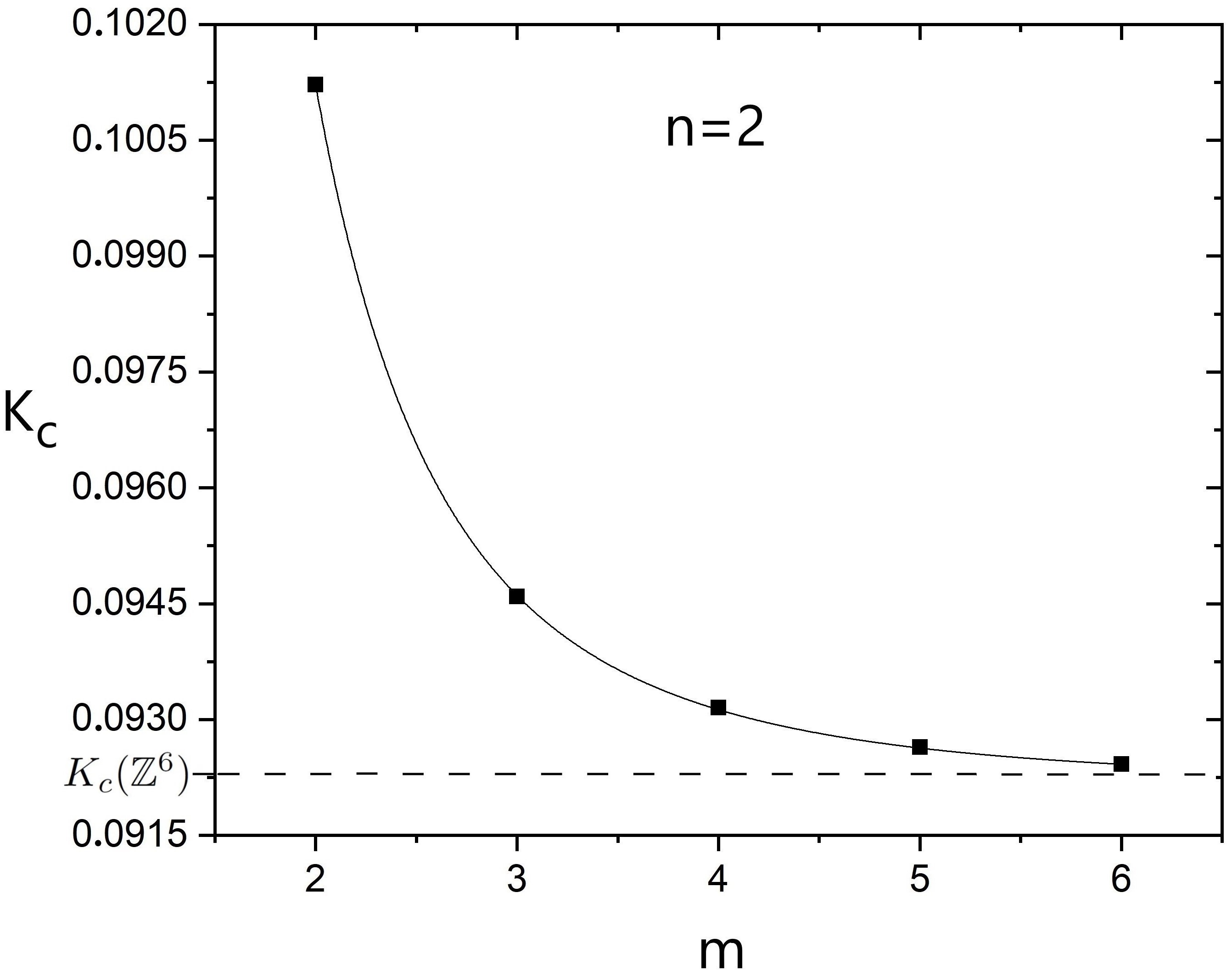}
		\caption{Analysis of $ K_c (m) $, estimated through $\chi$, as a function of $ m $. Left: $ n = 1 $, right: $ n = 2 $.
		The inverse critical temperature appears to converge monotonically to $ K_c (\mathbb{Z}^{4})=0.1496947(5) $ \cite{lundow} ($ n = 1 $) and to $ K_c(\mathbb{Z}^{6})=0.09229(4)$ \cite{aktekin1} ($ n = 2 $). The error bars are smaller than the symbols. The curves were obtained using the three-parameter fit (4).} \label{fig_reduz}
	\end{center}
\end{figure}

\begin{table*}[t]
	\centering
	\begin{small}
		\caption*{\ \textbf{Table 2:} Estimated inverse critical temperature for $n=1$ and $n=2$.}
		\begin{tabular}{c|ccc||c|ccccc}
			\hline
			$n$ & $m_{1}$    & \  \   $K_c$ ($\chi$) & \   \  $K_c$ $\left(\frac{\mbox{d}U}{\mbox{d}K}\right)$ \ \   & $n$  & $m_{1}$    & $m_2$  \ & \    \ $K_c$ ($\chi$) & \    \ $K_c$ $\left(\frac{\mbox{d}U}{\mbox{d}K}\right)$  \\  \hline
			\multirow{7}{*}{$1$} & 2      & \ \      0.173815(21)   &  \ \     0.173814(24) \ \   & \multirow{7}{*}{$2$}  & \multirow{1.2}{*}{2}    & \multirow{1.2}{*}{4} \   & \ \      \multirow{1.2}{*}{0.101255(16)}   &  \   \   \multirow{1.2}{*}{0.101218(22)}     \\
			& 5   & \ \    0.153280(23)      & \   \  0.153295(26) \  \  &  & \multirow{2.1}{*}{3}     & \multirow{2.1}{*}{9}  \    &   \ \   \multirow{2.1}{*}{0.094619(29)}    &  \  \  \multirow{2.1}{*}{0.094591(32)}     \\  
			& 8   &  \ \     0.150995(25)   & \ \    0.151095(29) \ \   &  & \multirow{3.1}{*}{4}    & \multirow{3.1}{*}{16}  \   &  \  \   \multirow{3.1}{*}{0.093133(42)}    & \ \    \multirow{3.1}{*}{0.093156(42)}    \\
			& 10  &  \ \    0.150501(28)    & \ \    0.150528(31)  \  \  &  &  \multirow{4.1}{*}{5}     & \multirow{4.1}{*}{25}   \  & \   \   \multirow{4.1}{*}{0.092596(54) }  & \   \  \multirow{4.1}{*}{0.092642(52)}     \\
			& 13  & \ \    0.150185(30)    & \ \    0.150139(33) \ \   & & \multirow{5.0}{*}{6}     & \multirow{5.0}{*}{36} \   & \ \   \multirow{5.0}{*}{0.092484(67)}    & \ \   \multirow{5.0}{*}{0.092397(62)}    \\
			& 16  & \ \     0.149941(32)  &      0.149975(36)   & \\
			& 19  & \ \     0.149888(34)   &       0.149896(38)  & \\
			\hline
		\end{tabular}
	\end{small}
\end{table*}

\begin{figure}[t]
	\begin{center}
		\vskip 0.1cm
		\includegraphics[height=6.5cm]{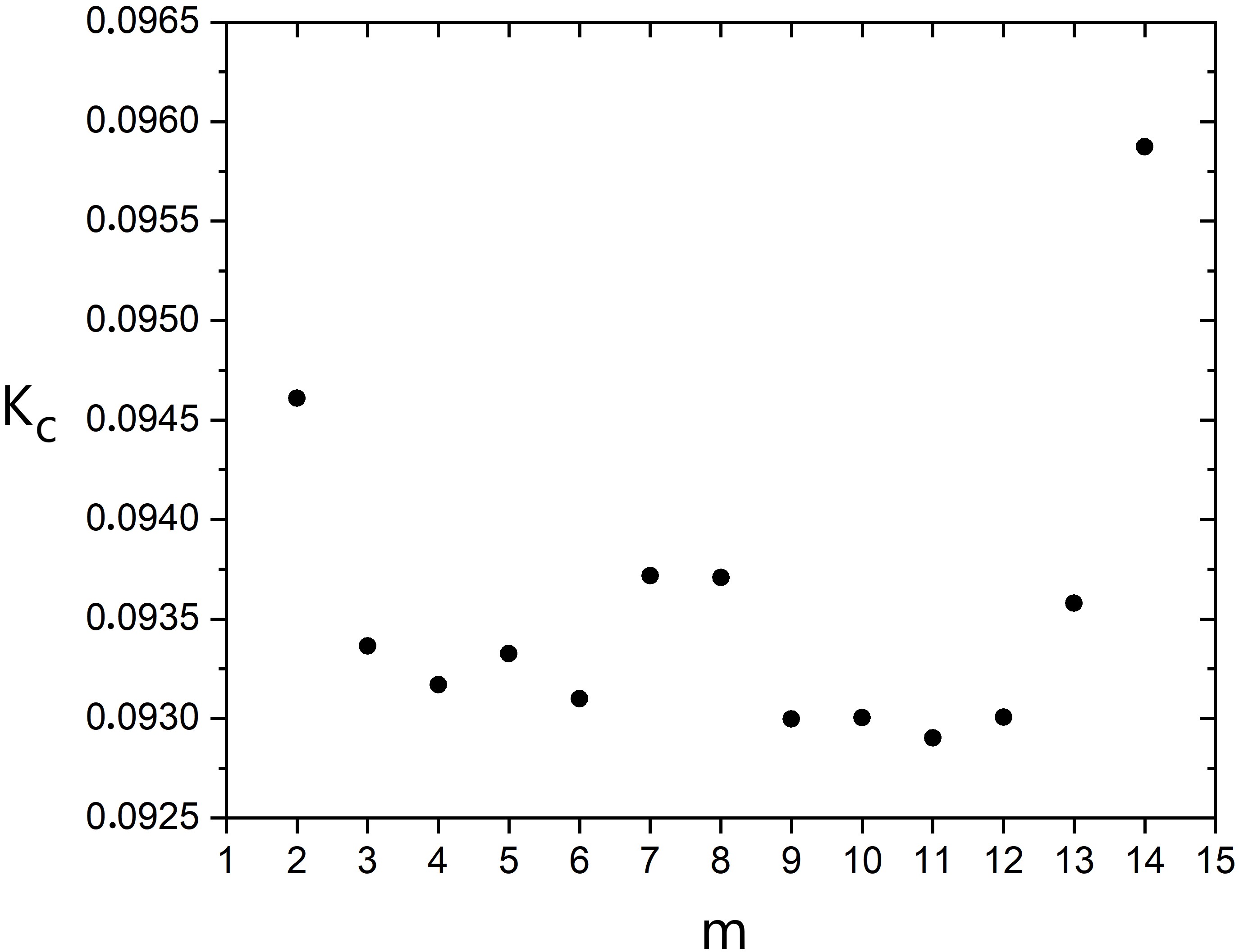}
		\caption{Effective inverse critical temperature $K_c(m; L)$, estimated through $\chi$, in model with sizes $1$, $m$, and $u$, where $u$ is fixed and $m$ varies between $1$ and $u$ ($L=1536$). The error bars are smaller than the symbols.} \label{irregular}
	\end{center}
\end{figure}

For the multirange percolation model with three ranges, $ 1 $, $ m $, and $ u $, the effective critical point varies in an irregular manner when $u$ is fixed and $ m $ varies between $ 1 $ and $ u $ \cite{Amaral}. We obtain numerical evidence showing analogous behavior in the multirange Ising model. The effective inverse critical temperature $ K_{c} (m; L) $ was estimated considering $ u = 15 $ and $1 < m < u$ for $L = 1536 $ using five independent samples (see Fig. \ref{irregular}). 

\section{Conclusion}

We study the two-dimensional Ising model with multiple interaction ranges. It is known that in percolation on $ \mathbb{Z}^ d$ with $n+1$ different ranges, each being a multiple of the previous one, the critical point converges to the critical point on $\mathbb{Z}^{d(n+1)}$ for $n \geq 1$ and $d\geq 2$. We show, via numerical simulation, that when we consider the critical temperature instead of the critical point the same result is valid for the Ising model if $ d=2 $ when $n=1$ or~$2$. We conjecture that the more general result valid for percolation is also valid for the Ising model.

For the case with three interaction ranges ($n=2$) we find that when the length of the largest range is fixed, then the critical temperature behavior is irregular if the range with intermediate length varies. This fact supports the hypothesis that the length of each range has to be a multiple of the one of the length immediately below.

The present study raises the question whether other models with local interactions and which exhibit phase transitions have similar connections between the number of interaction ranges and the critical temperature. In addition, it also allows us to estimate the critical temperature for the higher-dimensional Ising models by simulating the multirange Ising model on $ \mathbb{Z}^2 $, raising the possibility of a computationally efficient method to study critical properties of models in higher dimensions.



\section*{Acknowledgements}

We acknowledge the referees for their useful corrections and suggestions on the study. We thank L. Turban for helpful discussions. The authors would like to thank  the Brazilian agencies FAPEMIG, CAPES, and CNPq for their financial support.
 R. D. thanks CNPq for financial support under Project No. 303766/2016-6. 
 B. N. B. L. and A. P. F. A. thank CNPq for their financial support under Grants No.  305811/2018-5 and No. 308792/2018-1, respectively.

\vskip 1.5cm

\section*{Appendix: Uncertainty analysis.}

Due to the large amount of data analyzed and the extensive time required to obtain it, we have simplified obtaining uncertainties. To explain the methodology, we will separate the cases analyzed into two groups. Group I consists of cases $ m = 2 $ and $ m = 19 $ for $ n = 1 $, and $ m = 2 $ and $ m = 6 $ for $ n = 2 $. Group II contains the remaining cases. We describe the steps used to estimate the uncertainty of each parameter in the following.

\vskip 0.8cm

\noindent (1) {\it Uncertainty in $K_c(m;L)$ - Group I}

\vskip 0.3cm

For each $n$, $m$, and $L$ in this group, we obtain five independent samples of the functions $\chi$ and $U$ for $q$ values of the inverse temperature in the vicinity of $K_c(m; L)$ ($q=10$).

Let $K_{c, \chi}^{(i)} (m; L)$ be the value of $K$ that maximizes $\chi$ for a given $m$ and $L$ in the $i$-$th$ sample. $K_{c, \chi}^{(i)} (m; L)$ is determined by fitting a cubic spline to the simulation data ($q$ points) and locating the maximum of this polynomial. We estimate the uncertainty in $K_{c, \chi} (m; L)$ as the standard deviation of the mean. 

The estimates for $K_{c, U'}(m;L)$ and the uncertainties obtained by maximizing the derivative of $U$ are derived in a similar manner. The derivative is determined fitting $U$ using a five-parameter logistic function $U = a_1 + \frac{a_2 - a_1}{(1 + (a_3/K)^{a_4})^{a_5}}$ to the data for $U$ in the vicinity of $K_c(m,L)$, again using $q$ points. We then calculate the derivative of the logistic function to estimate $ \frac {\mbox{d}U} {\mbox{d}K} (U') $.

\vskip 0.8cm

\noindent (2) {\it Uncertainty in $K_c(m)$ - Group I}

\vskip 0.3cm

We estimate the critical exponent $\nu$ using the relation (2) and, using the relation (3), perform a three-parameter fit to determine the central value $K_c(m)$ and a fitting uncertainty $\Delta_f$. We calculated $\delta_{L_i}$, the uncertainty in $K_c(m)$ induced by $\Delta K_{c, \bullet} (m; L_i)$, as the difference in the value obtained through the fit of points using $K_{c,\bullet}(m; L_i) + \Delta K_{c,\bullet} (m; L_i)$ and $K_c(m)$, with all other $K_{c,\bullet}(m; L_i)$'s taking their central values (the symbol $\bullet$ denotes $\chi$ or $U'$).

The final uncertainty estimate in $K_c(m)$ is

\begin{equation*}
\Delta K_{c} (m) = \sqrt{(\Delta_f )^2 + \sum_{i=1}^{s} \delta_{L_i}^2}
\tag{6}
\label{three_p}
\end{equation*}

\noindent where $s$ denotes the number of $L$'s analyzed.

\vskip 0.8cm

\noindent (3) {\it Uncertainty in $K_c(m)$ - Group II}

\vskip 0.3cm

We did not estimate uncertainties for $ K_{c,\bullet}(m; L) $ in this group, as these data were obtained in single runs. To estimate the uncertainty in $ K_c (m) $ we perform a linear interpolation using the uncertainties obtained for group I (Table 3). We assume that $\Delta K_c(m)$ grows with $m$ due to finite size effects; the uncertainties determined for group I support this assumption.

\vskip 0.2cm

\begin{table*}[h]
	\centering
	\begin{small}
		\caption*{\ \textbf{Table 3:} Uncertainty estimates for group I.}
		
		\begin{tabular}{c|c|cc}
			\noalign{\smallskip}
			\hline
			\ $n$ \ \ & \  $m$  \ & \ \ \ \ \  $\Delta K_{c}(m) (\chi)$  & \ \ \ \ \  $\Delta K_{c}(m) (U')$  \\
			\hline
			\multirow{2}{*}{$1$} \   &  \ $2$ \ & \ \ \ \ \  $2.13 \times 10^{-5}$  & \ \ \ \ \  $2.37 \times 10^{-5}$  \\
			& \ $19$ \ & \ \ \ \ \ $3.35\times 10^{-5}$  & \ \ \ \ \  $3.77 \times 10^{-5}$   \\
			\hline
			\multirow{2}{*}{$2$} \ \ &  \ $2$ \  &  \ \ \ \ \   $1.62 \times 10^{-5}$     & \ \ \ \ \  $2.21 \times 10^{-5}$ \\
			&  \ $6$  \  &  \ \ \ \ \   $6.68 \times 10^{-5}$    & \ \ \ \ \  $6.20 \times 10^{-5}$    \\
			\hline
		\end{tabular}
	\end{small}
\end{table*}

\vskip 0.8cm

\noindent (4) {\it Uncertainties in $K_c(\mathbb{Z}^4)$ and $K_c(\mathbb{Z}^6)$}

\vskip 0.3cm

For estimating the uncertainties in $K_c(\mathbb{Z}^4)$ and $K_c(\mathbb{Z}^6)$ via Eq. (4), we use the same procedure used to estimate $ \Delta K_{c}(m)$; in this case we have $q=7$ data points for $ n = 1 $, and $ q = 5 $ for $ n = 2 $ ($q$ is the number of $ m $ values analyzed).

\end{document}